\documentclass[preprint,aps,prd,showpacs,floatfix]{revtex4-1}
\usepackage{times}
\usepackage {euscript}
\usepackage{amsthm}
\usepackage{graphicx}
\usepackage{amsmath}
\usepackage{amssymb}
\usepackage{amsfonts}
\usepackage[english]{babel}
\usepackage{epstopdf}
\usepackage{multirow,makecell,array}
\usepackage{mathtools}
\usepackage{epstopdf}
\usepackage{color}
\begin{document}
\thispagestyle{empty}

    \title{Pulse shape effects on the electron-positron\\pair production in strong laser fields}

\author{I.~A.~Aleksandrov$^{1, 2}$}\author{G.~Plunien$^{3}$} \author{V.~M.~Shabaev$^{1}$}
\affiliation{$^1$~Department of Physics, St. Petersburg State University, 7/9 Universitetskaya nab., Saint Petersburg 199034, Russia\\$^2$~ITMO University, Kronverkskii ave 49, Saint Petersburg 197101, Russia\\$^3$~Institut f\"ur Theoretische Physik, TU Dresden, Mommsenstrasse 13, Dresden, D-01062, Germany
\vspace{10mm}
}

\begin{abstract}
The pair-production process in the presence of strong linearly polarized laser fields with a subcycle structure is considered. Laser pulses with different envelope shapes are examined by means of a nonperturbative numerical approach. As generic cases, we analyze two different ``flat'' envelope shapes and two shapes without a plateau as a function of various characteristic parameters including the carrier-envelope phase, respectively. The resonant Rabi oscillations, the momentum distribution of particles created, and the total number of pairs are studied. It is demonstrated that all these characteristics are very sensitive to the pulse shape.
\end{abstract}

\pacs{12.20.-m, 12.20.Ds, 11.15.Tk, 42.50.Xa}

\maketitle
\section{Introduction}\label{sec:intro}
\indent The phenomenon of particle-antiparticle production in the presence of strong external electromagnetic fields has been a focus of numerous theoretical investigations since the 1930s~\cite{sauter_1931, euler_heisenberg}. However, it has been observed only within the multiphoton regime~\cite{burke_prl_1997}, where perturbation theory can be successfully employed. In the nonperturbative regime this (Schwinger) effect~\cite{schwinger_1951} has never been confirmed by experiments [the characteristic critical field strength is $E_\text{c} = m^2 c^3/(|e| \hbar) \approx 1.3 \times 10^{16}$ V/cm, where $m$ and $e$ are the electron mass and charge, respectively]. In this context various scenarios involving laser fields seem to be rather promising. Although a rigorous theoretical formalism based on quantization of the Dirac field in the presence of an external electromagnetic background was thoroughly elaborated (see~\cite{fradkin_gitman_shvartsman} and references therein), many aspects of the $e^+e^-$ pair-production phenomenon in strong laser fields still need to be elucidated. The present investigation aims at the numerical study of the pulse shape effects on the pair-creation process.

\indent Since the exact solutions of the Dirac equation are found analytically for a very few configurations of the external background, a detailed analysis can only be provided by means of  corresponding numerical calculations. We perform a numerical integration of the Dirac equation taking into account the interaction with the external laser field (i.e. nonperturbatively). The pair-production probabilities are evaluated with the aid of the general formalism described in Ref.~\cite{fradkin_gitman_shvartsman} (see also Refs.~\cite{gav_git_prd_1996, adorno_phys_scr_2015, adorno_2015, adorno_2016}).

\indent If one introduces the adiabaticity parameter $\xi = |eE_0|/(mc\omega)$, where $E_0$ is the peak electric field strength and $\omega$ is the carrier frequency, then the tunnelling and multiphoton regimes can be characterized by $\xi \gg 1$ and $\xi \ll 1$, respectively~\cite{keldysh} (the parameter $\xi$ is the inverse of the Keldysh parameter $\gamma$~\cite{keldysh}). In the present study we focus on the intermediate case $\xi = 1$, when the process exhibits a multiphoton behavior and, at the same time, reflects a nonperturbative nature.

\indent A number of previous studies~\cite{heinzl_plb_2010, linder_prd_2015, hebenstreit_prl_2009, abdukerim_plb_2013, dumlu_dunne_prd_2011, hebenstreit_prd_2010, jansen_prd_2016, hebenstreit_plb_2014, nuriman_plb_2012, kohlfuerst_prd_2013, titov_prd_2016} indicated the importance of the pulse shape and carrier-envelope phase (CEP) effects on the pair-production process in various conditions. This motivated us to further scrutinize these effects regarding the case of a strong individual laser pulse having a temporal dependence (the field is assumed to be spatially homogeneous). Such a configuration approximates the experimental scenario of two counterpropagating linearly polarized laser pulses.

\indent We consider two types of temporal profiles. The type-I pulses are characterized by a wide plateau, where the pulse envelope equals unity. The pulses of the type II have no plateau and their envelopes immediately start to diminish once they reach the maximal value of unity. One of the type-I pulse shapes was examined in detail in Ref.~\cite{mocken_pra_2010}, however, in this analysis only the pulse duration was taken as varying parameter, while the effects of the envelope shape and CEP were not discussed. We shall study two different profiles of the type I being characterized by three independent parameters: the pulse duration, the duration of switching on and off (rapidity of the pulse switching), and CEP. It will be demonstrated that each of these parameters may have a notable impact on the patterns established previously. Besides, we examine two different profiles of type II: $\cos^2$ and Sauter-like pulses. In this case one can separately vary the pulse duration and CEP. It will be shown that the main features of the pair-production process in the case of the type-II pulse shapes considerably differ from those for the type-I case.

\indent In particular, the comparison of these four pulse shapes allows one to reveal several important features of the resonant Rabi pattern, momentum spectrum of particles created, and total number of pairs as a function of the laser frequency and pulse duration. Some of these properties were already discussed in the literature~\cite{abdukerim_plb_2013, hebenstreit_prl_2009}. In Ref.~\cite{hebenstreit_prl_2009} a number of characteristic signatures in the spectrum of particles produced were found (only Gaussian envelope was analyzed). In Ref.~\cite{abdukerim_plb_2013} various Gaussian and super-Gaussian envelope shapes were examined. We extend these studies by
providing results based on an independent numerical approach. We consider the present investigation of the pulse shape effects to be particularly comprehensive for two reasons. First, we provide the analysis of all the important characteristics of the pair-production process (listed above). Second, we study a broad class of possible pulse shapes.
%This allows one to further clarify many aspects of the pair-production phenomenon in the presence of strong laser fields.

\indent The paper is organized as follows. In Sec.~\ref{sec:env_shape} we describe the temporal profiles to be analyzed. In Secs.~\ref{sec:type-I} and \ref{sec:type-II} we discuss the pulse shape effects for type-I and type-II envelopes, respectively. In Sec.~\ref{sec:total} we present results 
for the total number of pairs produced for various envelope shapes, laser frequencies, and pulse durations, respectively. Sec.~\ref{sec:cep} contains the analysis of the CEP effects regarding both type-I and type-II pulses. Finally, in Sec.~\ref{sec:discussion} we summarize our results and provide a discussion.

\indent Relativistic units ($\hbar = 1$, $c = 1$) together with the Heaviside charge unit ($\alpha = e^2/4\pi$, $e<0$) are employed throughout the paper. Accordingly, the Schwinger critical field strength is $|e| E_\text{c} = m^2$, one relativistic unit of length is $\hbar/(mc) \approx 3.862 \times 10^{-13}~\text{m}$, and the unit of time is $\hbar/(mc^2) \approx 1.288 \times 10^{-21}~\text{s}$.
%
%%%
\section{Envelope shapes to be examined}\label{sec:env_shape}
\begin{figure}[h]
\center{\includegraphics[height=0.34\linewidth]{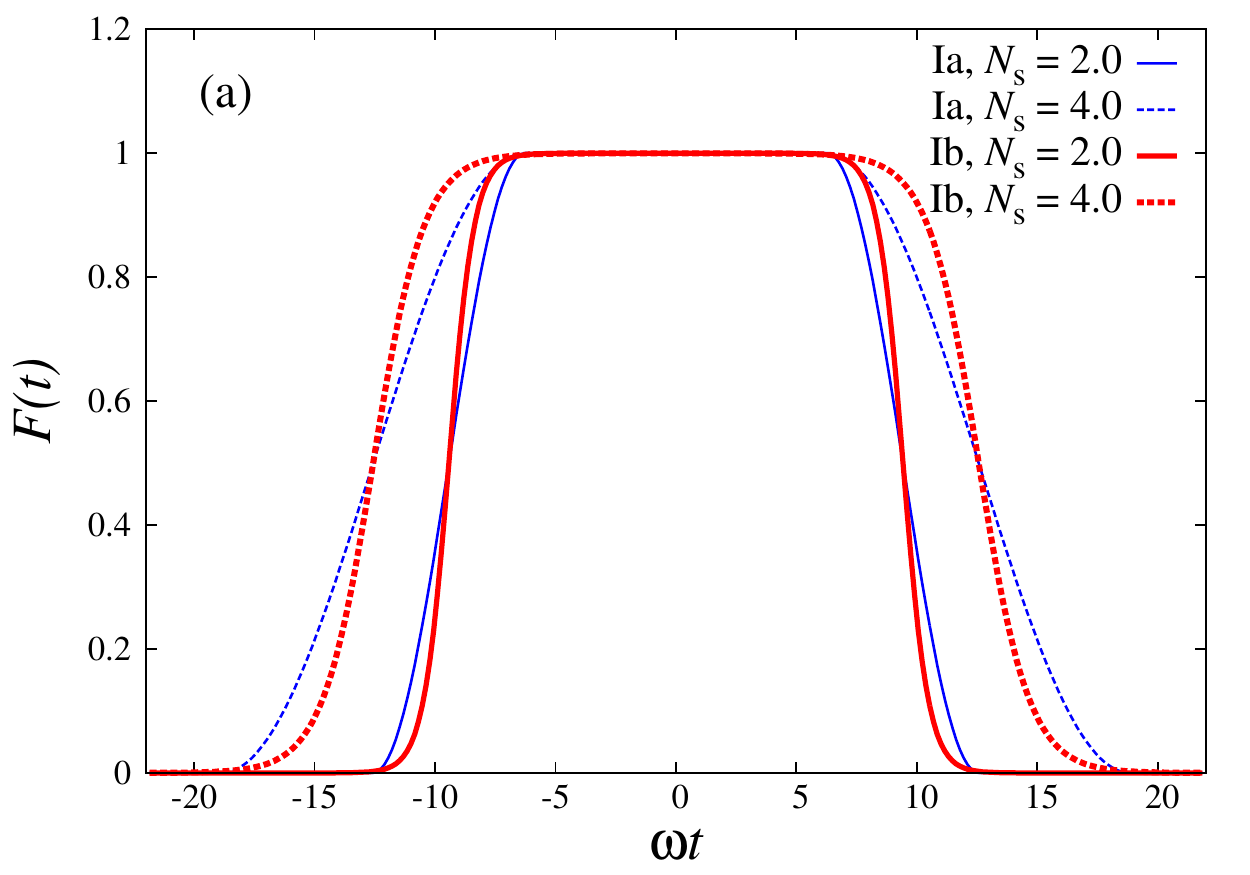}~
\includegraphics[height=0.34\linewidth]{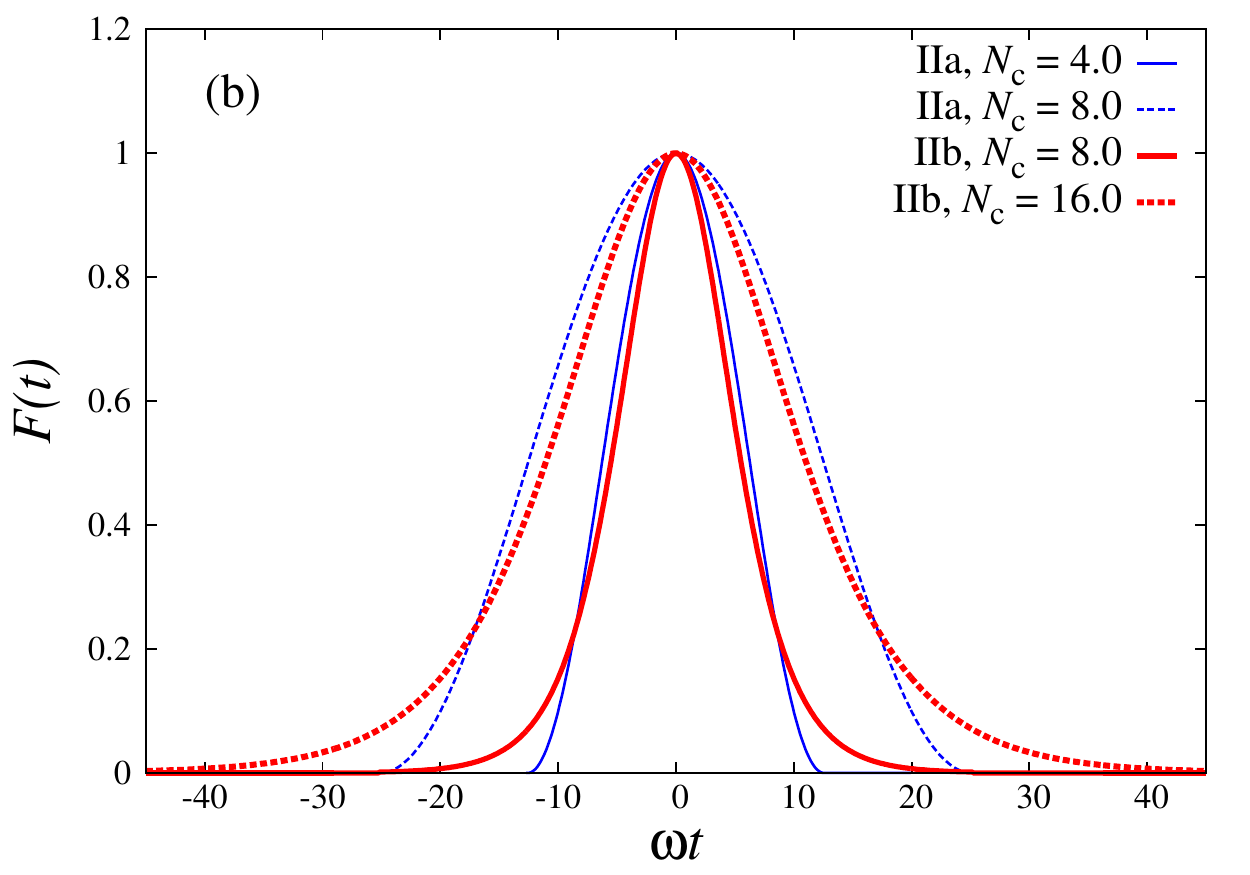}}
\caption{(a)~Type-I envelope shapes for $N_\text{c} = 2.0$ and two different values of $N_\text{s}$, (b)~Type-II envelope shapes for various $N_\text{c}$.}
\label{fig:shapes}
\end{figure}

%%%
We assume that the external electric field has the following form:
\begin{equation}
E_z (t) = E(t) = E_0 F(t) \sin (\omega t + \varphi)\,,\quad E_x = E_y = 0\,,\label{eq:field_gen}
\end{equation}
where $E_0$ is the peak electric field strength, $F(t)$ is a smooth envelope function ($0 \leq F(t) \leq 1$), $\omega$ is the carrier frequency, and $\varphi$ describes CEP. First, we consider $\varphi = 0$ (the CEP effects will be mainly discussed in Sec.~\ref{sec:cep}). Within the present study we analyze two envelope profiles having an extended plateau region (type I) and two forms of an envelope without a plateau (type II). Namely, they have the following expressions:
\begin{eqnarray}
F_\text{Ia} (t) &=& 
\begin{cases}
\sin^2 \frac{\pi (- T/2 + \Delta T - t)}{2 \Delta T} &\text{if}~~ - T/2 -\Delta T \leq t \leq -T/2,\\
1 &\text{if}~~-T/2 \leq t \leq T/2\,,\\
\sin^2 \frac{\pi (T/2 + \Delta T - t)}{2 \Delta T} &\text{if}~~T/2 \leq t \leq T/2 + \Delta T\,,\\
0 &\text{otherwise},
\end{cases} \label{eq:F_Ia}\\
F_\text{Ib} (t) &=& \frac{1}{2} \bigg [ \tanh \frac{6 (t+T/2+\Delta T/2)}{\Delta T} - \tanh \frac{6 (t-T/2-\Delta T/2)}{\Delta T} \bigg ], \label{eq:F_Ib}\\
F_\text{IIa} (t) &=& \cos^2 \bigg ( \pi \frac{t}{T} \bigg ) \, \theta (T/2 - |t|), \label{eq:F_IIa}\\
F_\text{IIb} (t) &=& \frac{1}{\cosh^2 (8t/T)}\,. \label{eq:F_IIb}
\end{eqnarray}
The external laser pulse contains $N_\text{c} = \omega T/(2\pi)$ cycles of the carrier. For the type-I pulses the switching part contains $N_\text{s} = \omega \Delta T /\pi$ half cycles. The factors $6$ and $8$ in Eqs.~(\ref{eq:F_Ib}) and (\ref{eq:F_IIb}), respectively, guarantee that the values of $\Delta T$ and $T$ correspond to the duration of the switching part and the total pulse duration, respectively (in fact, they can slightly differ from $6$ and $8$). In order to make sure that $\varphi$ is the only parameter responsible for the CEP effects, we always choose $N_\text{c}$ so that $N_\text{c} + N_\text{s} = 2k$ ($k \in \mathbb{N}$). This leads to
\begin{equation}
\sin(\omega t + \varphi) \Big |_{t=t_\text{in}=-T/2-\Delta T} = \sin \varphi\,.
\label{eq:cep_in}
\end{equation}
Thus, $\varphi$ is the carrier phase at the initial time instant $t_\text{in}$. In fact, this specific choice of $N_\text{c}$ does not impose any significant restrictions on our computations, while the symmetry of the envelope $F(t)$ becomes useful for the necessary derivations. In Fig.~\ref{fig:shapes}(a) we display the Ia and Ib envelopes for $N_\text{c} = 2.0$ and two different values of $N_\text{s}$. In Fig.~\ref{fig:shapes}(b) the profiles IIa and IIb are depicted for various $N_\text{c}$.
Note that these envelopes possess the following properties:
\begin{eqnarray}
\int \limits_{-\infty}^{+\infty} F_\text{Ia} (t) \mathrm{d}t &=&\int \limits_{-\infty}^{+\infty} F_\text{Ib} (t) \mathrm{d}t = T + \Delta T, \label{eq:F_int_I}\\
\int \limits_{-\infty}^{+\infty} F_\text{IIa} (t) \mathrm{d}t &=& 2 \int \limits_{-\infty}^{+\infty} F_\text{IIb} (t) \mathrm{d}t = T/2\,. \label{eq:F_int_II}
\end{eqnarray}

\indent In the next section the type-I envelope shapes of the external pulse will be analyzed.
%%%
\section{Type-I envelope shapes}\label{sec:type-I}
%%%
Solving the Dirac equation in the corresponding external potential on a temporal grid, we evaluate all the necessary pair-production probabilities using the formalism of the in and out complete sets of time-dependent solutions~\cite{fradkin_gitman_shvartsman} (see Appendix~\ref{sec:appendix}). All of the envelope shapes described above are analyzed with the help of the same numerical procedures.

\indent Let us introduce a number density of electrons produced per unit volume in momentum space:
\begin{equation}
n (\boldsymbol{p}) = \frac{(2\pi)^3}{V}\frac{\mathrm{d} N_{\boldsymbol{p}, r}^\text{(el)}}{\mathrm{d}^3 \boldsymbol{p}}\,,
\label{eq:n_notation}
\end{equation}
where $V$ is the volume of the system and $r=\pm 1$ denotes the electron spin state \big [in fact, $n(\boldsymbol{p})$ is independent of $r$\big ]. First, we consider the number density of electrons created at rest $n (\boldsymbol{p} = 0)$. For the case of the type-I envelope shapes this quantity oscillates as a function of $N_\text{c}$ (see, e.g., Refs.~\cite{popov_1973, narozhniy_1974, avetissian_pre_2002, mocken_pra_2010}). We start with the analysis of the corresponding Rabi oscillations for $\xi=1.0$. As will be shown, there are several pronounced differences between the Ia and Ib pulse shapes regarding this aspect.
%%%
\subsection{Resonant Rabi oscillations}\label{subsec:I_rabi}
\begin{figure}[h]
\center{\includegraphics[height=0.31\linewidth]{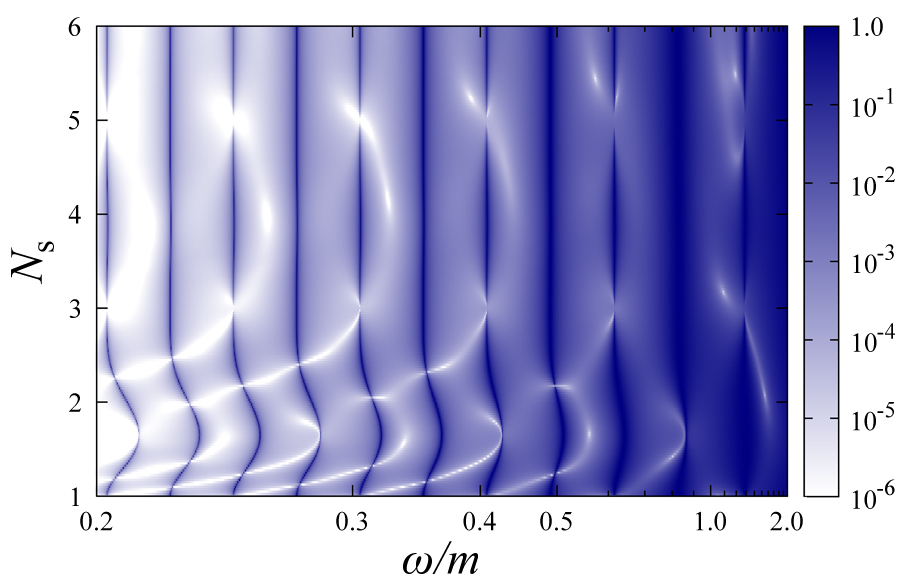}~~
\includegraphics[height=0.31\linewidth]{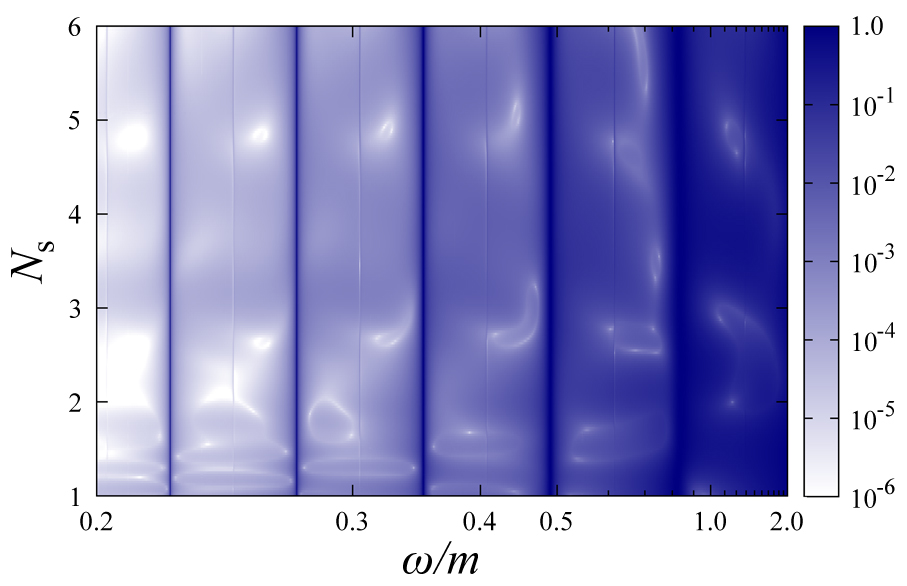}}
\caption{Resonant picture as a function of $N_\text{s}$ for the Ia (left) and Ib (right) envelope shapes. The integer numbers in the white boxes indicate the numbers $n$ for the corresponding $n$-photon peaks.}
\label{fig:res_I_to}
\end{figure}

%%%
First, we evaluate the maximal values of the function $n (\boldsymbol{p} = 0)$ (as a function of $N_\text{c}$ in the range $0< N_\text{c} \leq 400$) for various values of $\omega$ and $N_\text{s}$. In Fig.~\ref{fig:res_I_to} the corresponding dependences are presented for both Ia and Ib envelope shapes. For a given $N_\text{s}$ one observes a resonant pattern which consists of a set of equidistant peaks (with respect to the reciprocal $\omega$ axis). Each resonance relates to a certain number $n$ of laser photons required for pair production ($\omega_n = 2m_*/n$, where $n=1,2,3...$ and $m_*$ is the laser-dressed effective mass of electrons which is almost independent of $\omega$~\cite{mocken_pra_2010}). In the case of the Ia pulse the even resonances disappear for odd values of $N_\text{s}$. In the Ib case they are strongly suppressed for all $N_\text{s}$ and the picture is almost independent of this parameter. If $N_\text{s} \lesssim 2.5$, the positions of the resonant peaks in the Ia case depend on $N_\text{s}$, which means that the effective mass $m_*$ depends on this parameter.

\begin{figure}[h]
\center{\includegraphics[height=0.4\linewidth]{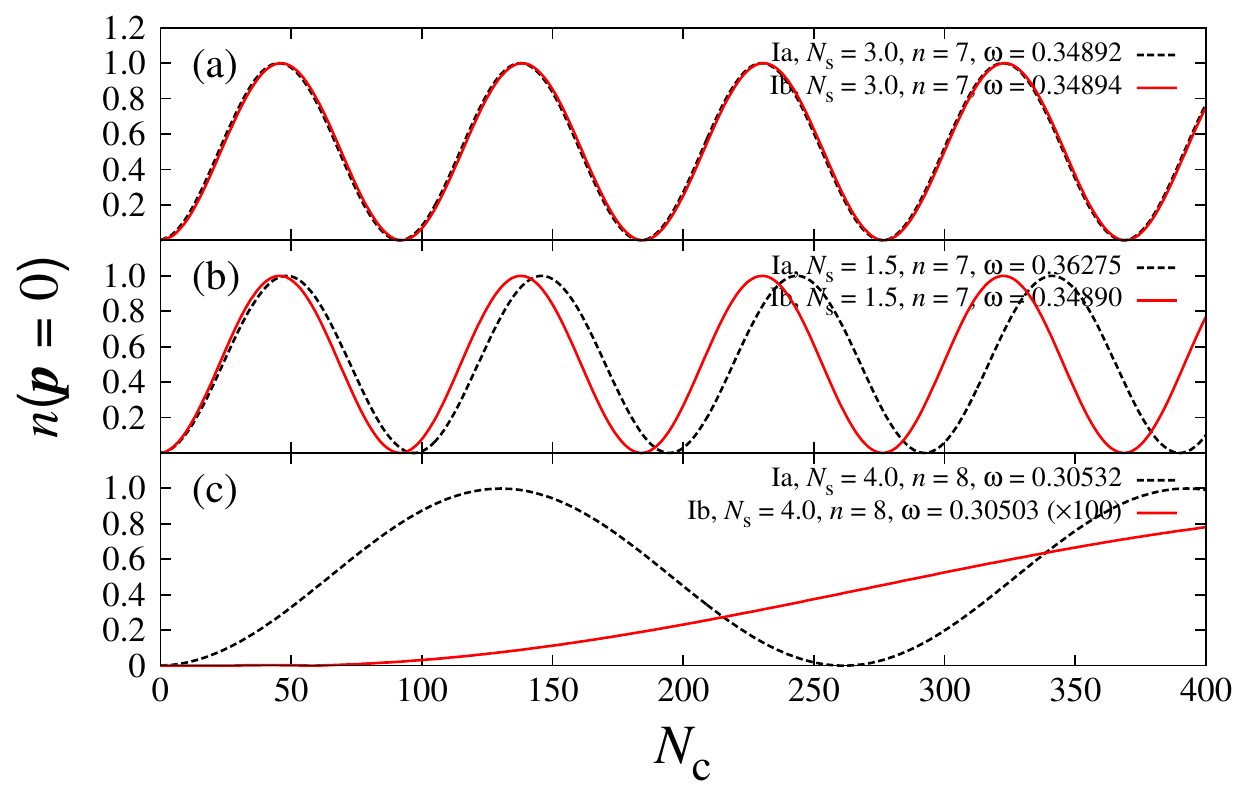}}
\caption{Resonant Rabi oscillations for: (a)~$n=7$ resonance for $N_\text{s} = 3.0$ and both Ia and Ib pulse shapes, (b)~$n=7$ resonance for $N_\text{s} = 3.0$, (c)~$n=8$ resonance for $N_\text{s} = 4.0$ (the values of the Ib case are multiplied by $100$).}
\label{fig:I_rabi}
\end{figure}

\indent In order to gain a better understanding of the features described above, we present the corresponding resonant Rabi oscillations for the $n=7$ and $n=8$ resonances and various $N_\text{s}$ (see Fig.~\ref{fig:I_rabi}).
While for the odd resonances in the region $N_\text{s} \gtrsim 2.5$ the oscillations for the Ia and Ib shapes are identical [Fig.~\ref{fig:I_rabi}(a)], for smaller $N_\text{s}$ the Rabi frequencies are different [Fig.~\ref{fig:I_rabi}(b)]. Moreover, when one considers an even resonance, this difference becomes dramatic. The Rabi frequency in the case of the Ib envelope is extremely low [Fig.~\ref{fig:I_rabi}(c)], which does not allow the pair-production probability to reach its maximum due to the restriction $N_\text{c} \leq 400$. In fact, for the specific parameters employed in Fig.~\ref{fig:I_rabi}(c) this maximal value equals 0.00883 and appears at $N_\text{c} \approx 500$. In the Ia case the corresponding Rabi frequencies are much greater and the maxima equal unity. Thus, all the resonances in Fig.~\ref{fig:res_I_to}(left) are clearly seen.

\indent To our knowledge, the even peaks of a unit height have not been observed within the previous numerical studies. This can be explained by the charge-conjugation symmetry~\cite{mocken_pra_2010, akal_prd_2014}. For $\boldsymbol{p}=0$ the electron-positron pair must have an odd C parity and, therefore, it can be produced only via an absorption of odd number of photons (the C parity of a single photon is also odd). However, in the case of the Ia pulse shape this explanation fails. We suppose that the reason for this is that the external field is not monochromatic so it contains photons with various energy $k_0 \neq \omega$. For instance, in the presence of two monochromatic fields having the frequencies $\omega$ and $\omega' \ll \omega, m_*$ the resonant $(n+1)$-photon process which relates to $n\omega + \omega' = 2m_*$ is not prohibited for even $n$. Varying the frequency $\omega$, one finds additional ``even'' peaks at $\omega_n \approx 2m_* /2$, $2m_*/4$, and so on. This issue will be discussed in more detail in Sec.~\ref{sec:cep}.
%%%
\subsection{Momentum distribution}\label{subsec:I_mom}
\begin{figure}[h]
\center{\includegraphics[height=0.43\linewidth]{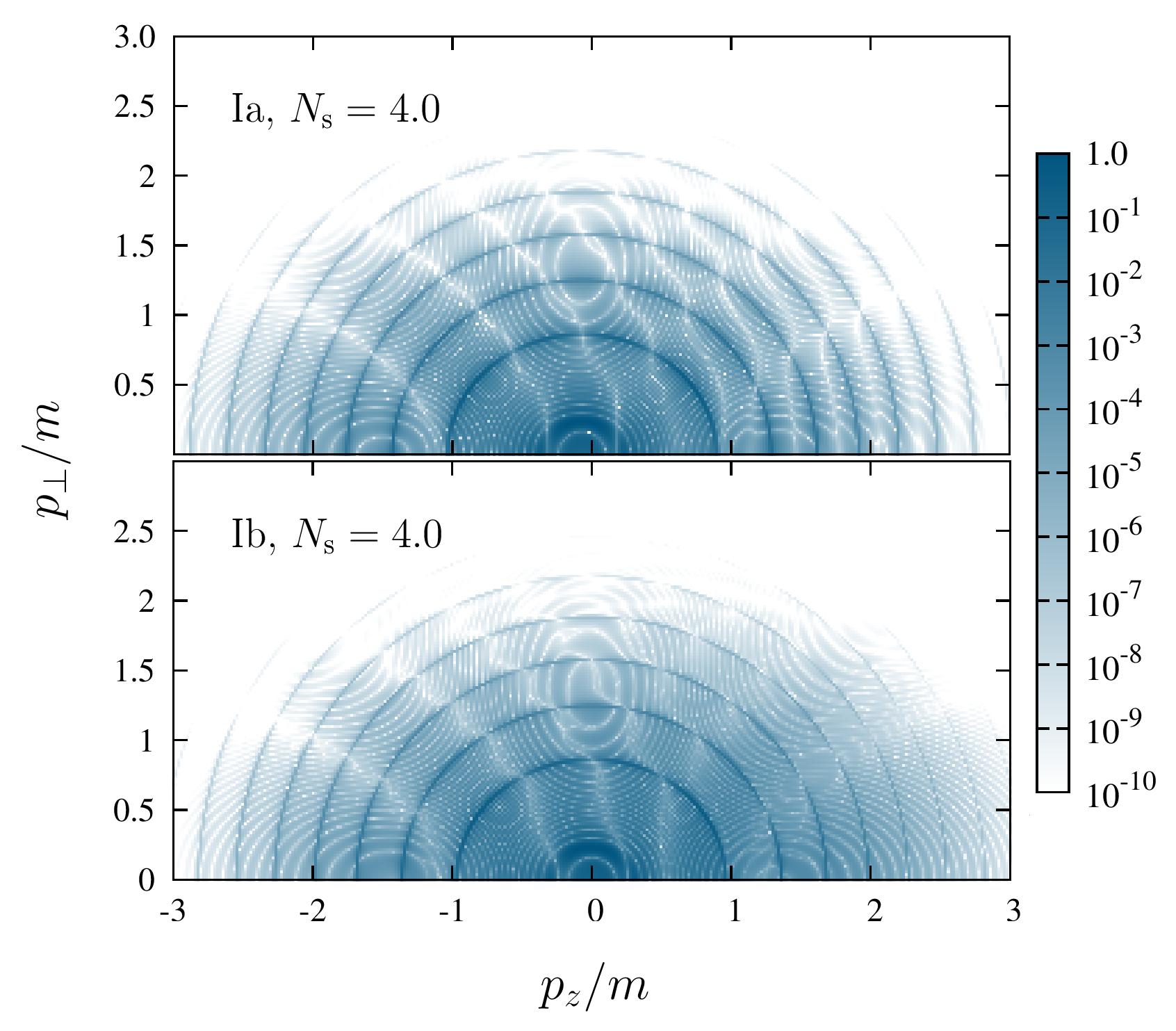}~
\includegraphics[height=0.43\linewidth]{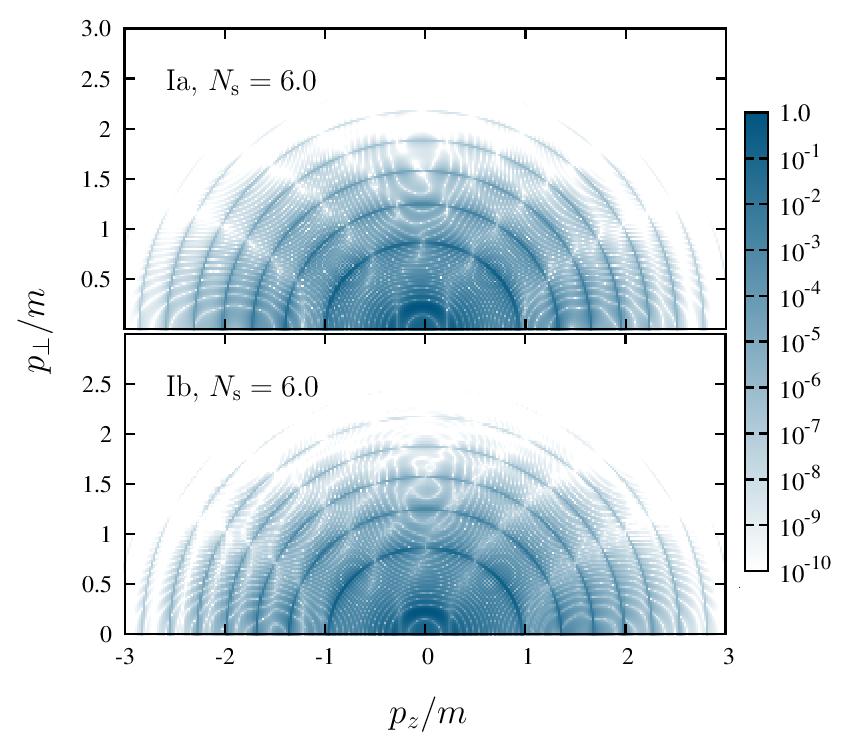}}
\caption{Momentum distribution of electrons created for $N_\text{c} + N_\text{s} = 20.0$, $\omega = 0.5m$, and both Ia and Ib pulse shapes for $N_\text{s} = 4.0$ (left) and $N_\text{s} = 6.0$ (right). The pulse duration is $T \approx 201 \, m^{-1}$ (left) and $T \approx 176 \, m^{-1}$ (right).}
\label{fig:I_mom}
\end{figure}

\begin{figure}[h]
\center{\includegraphics[height=0.243\linewidth]{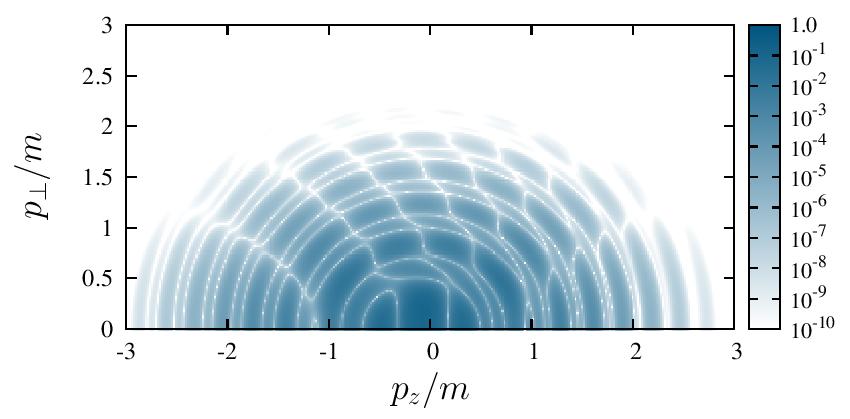}~
\includegraphics[height=0.243\linewidth]{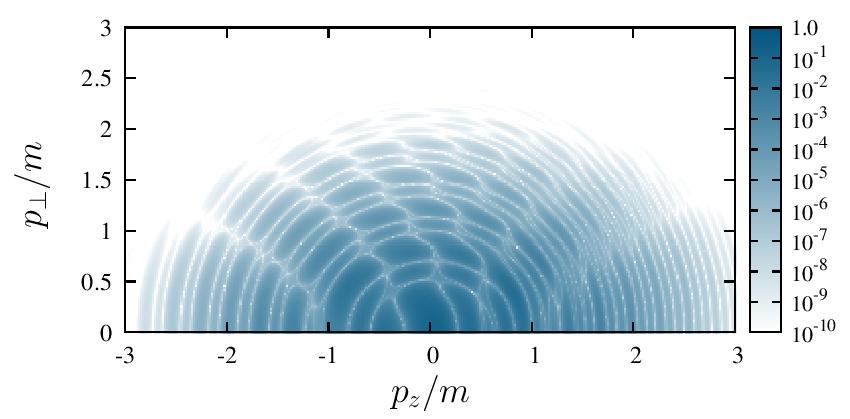}}
\caption{Momentum distribution of electrons created for $N_\text{c} = 2.0$, $N_\text{s} = 4.0$, $\omega = 0.5m$ ($T = \Delta T \approx 25.1 \, m^{-1}$), and both Ia (left) and Ib (right) envelope shapes.}
\label{fig:I_mom_short}
\end{figure}

%%%
We now turn to the discussion of the spectra of electrons created, i.e. we will consider the function $n(\boldsymbol{p})$ for various $\boldsymbol{p}$. In Fig.~\ref{fig:I_mom} the momentum distribution of electrons created is demonstrated for $N_\text{c} + N_\text{s} = 20.0$ (the pulse duration is fixed), $\omega = 0.5m$, two types of the envelope shape (Ia and Ib), and two different values of $N_\text{s}$. The electron produced has the longitudinal component of its momentum $p_z$ and the transverse component $p_\perp$. We vary both of them.
In Fig.~\ref{fig:I_mom} we observe a characteristic ring structure that can be accounted for by the same resonant condition $2m_* = n\omega$ where $m_*$ should be replaced by the laser-dressed effective energy which now depends on $\boldsymbol{p}$~\cite{mocken_pra_2010}. Each ring corresponds to a certain integer number $n$. Note that $p_z$ is the $z$ component of the gauge invariant momentum and the center of the momentum distribution does not necessarily coincide with the origin $\boldsymbol{p}=0$. We point out that the general structure of the momentum distribution is the same for all type-I pulses, whereas it could be different quantitatively. The pulse shape starts to play a more important role when the number of cycles in the pulse becomes smaller. For instance, in Fig.~\ref{fig:I_mom_short} we employ $N_\text{c} = 2.0$ and $N_\text{s} = 4.0$. Changing these parameters, one can produce a wide variety of distributions (for the case of a Gaussian envelope shape this was studied in Ref.~\cite{hebenstreit_prl_2009}).

\indent In this section we discussed various features of the pair-creation phenomenon. However, from the experimental viewpoint the most important characteristic of the process is the total number of pairs created. This will be discussed for both type-I and type-II pulse shapes in Sec.~\ref{sec:total}.
%%%
\section{Type-II envelope shapes}\label{sec:type-II}
\begin{figure}[h]
\center{\includegraphics[height=0.4\linewidth]{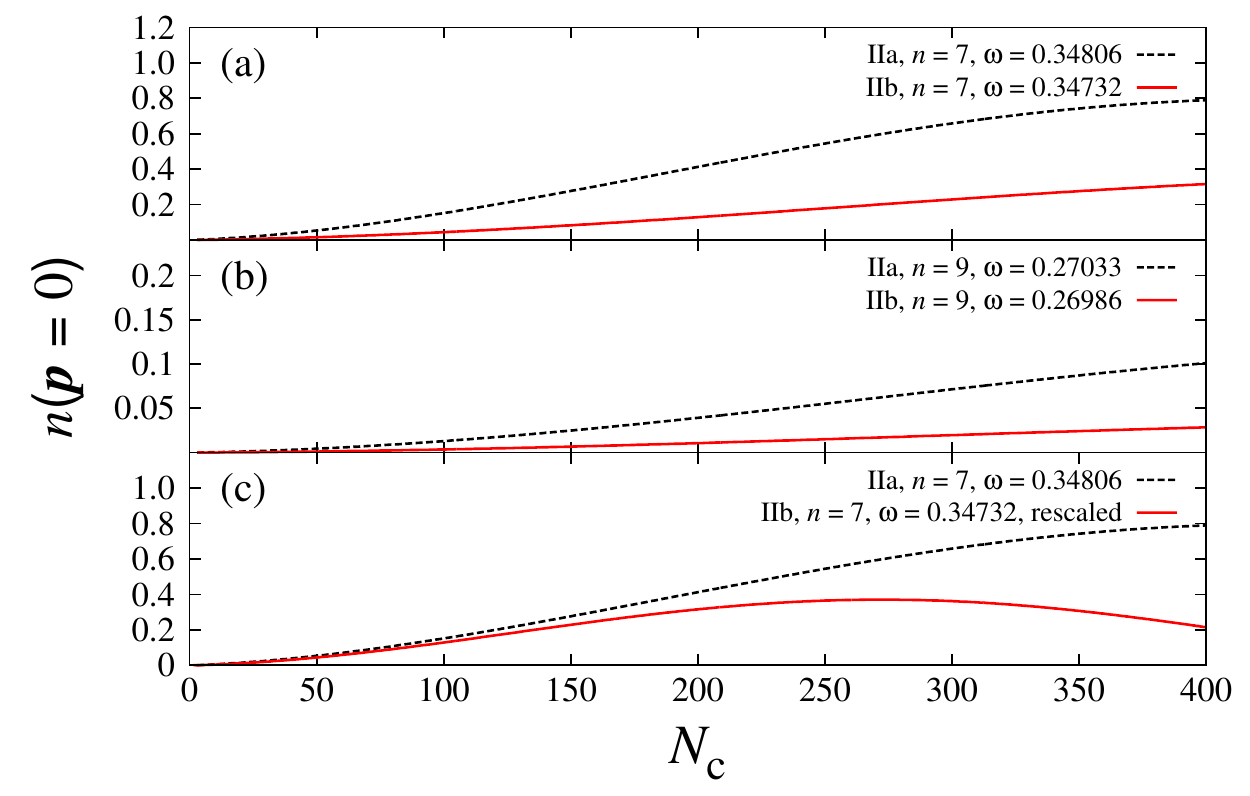}}
\caption{Resonant Rabi oscillations for: (a)~$n=7$ resonance for both IIa and IIb envelope shapes, (b)~$n=9$ resonance, (c)~$n=7$ resonance (for the IIb curve the bottom axis scale is chosen so that $N_\text{c} \in [0,~800]$).}
\label{fig:rabi_II}
\end{figure}

%%%
In contrast to the type-I envelope pulses, the type-II shapes do not have a plateau region and, therefore, the pulse duration and the way how it is switched on/off are now related. As a result, the Rabi oscillations are strongly modified. For the case of the type-II envelopes only odd resonances appear. In Fig.~\ref{fig:rabi_II} the corresponding Rabi oscillations are presented for various parameters of the external laser pulse. In contrast to the case of the type-I shapes, the pair-production probability now does not reach unity and its maximal value decreases with decreasing $\omega$ (i.e., for greater $n$). Moreover, the maximum for the IIa shape is much larger than that for the IIb shape (e.g., for the $n=7$ resonance they are $0.7977$ and $0.3706$, respectively). Besides, we note that the resonant Rabi frequencies in the type-II case are much smaller than those for the type-I pulses. In Fig.~\ref{fig:rabi_II}(c) we compare the Rabi oscillations for the IIa and IIb cases and display the latter rescaling the bottom axis by a factor of $2$ (i.e., the label $N_\text{c} = 400$ corresponds to $N_\text{c} = 800$ for the red solid line). We observe that this naive transformation, motivated by the property~(\ref{eq:F_int_II}), does not make the dependences in Fig.~\ref{fig:rabi_II}(c) coincide, which indicates a substantial difference between the IIa and IIb envelope shapes.

\begin{figure}[h]
\center{\includegraphics[height=0.24\linewidth]{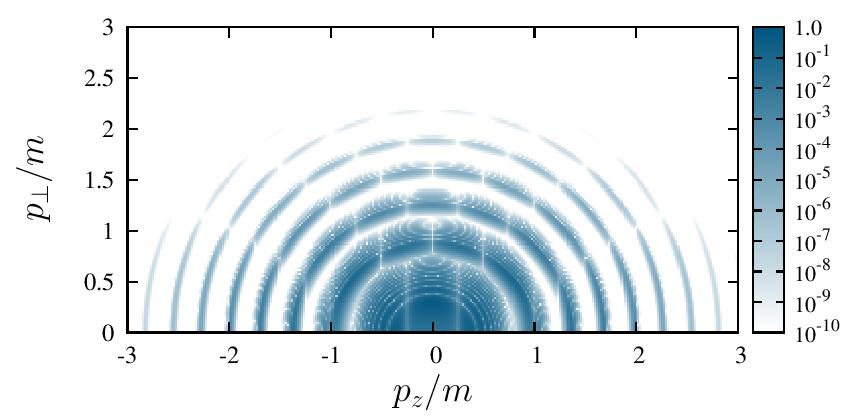}~
\includegraphics[height=0.24\linewidth]{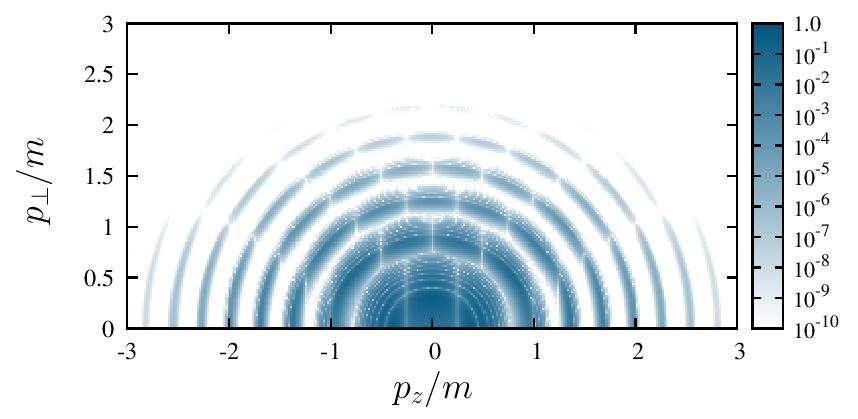}}
\caption{Momentum distribution of electrons created for $\omega = 0.5m$ and both IIa ($N_\text{c} = 40.0$, left) and IIb ($N_\text{c} = 80.0$, right) envelope shapes. The pulse duration is $T\approx 503 \, m^{-1}$ (left) and $T\approx 1005 \, m^{-1}$ (right).}
\label{fig:II_mom}
\end{figure}

\indent In Fig.~\ref{fig:II_mom} we display the momentum distribution of electrons created for the IIa pulse with $N_\text{c} = 40.0$ and the IIb pulse with $N_\text{c} = 80.0$.
These values of $N_\text{c}$ are chosen in order to keep the parameter $T_0 = \int F(t) \mathrm{d} t$ the same [see Eq.~(\ref{eq:F_int_II})]. The momentum distribution now considerably differs from that of the envelope shapes of type I (see Fig.~\ref{fig:I_mom}). In addition to the ring structure of the maxima, it contains a number of white rings regarding to extremely low probabilities. The distributions for the IIa and IIb envelope shapes are similar, provided the parameter $T_0$ is fixed. If one considers these two pulses of the same duration $T$, the results will be different (in the next section this will be discussed with respect to the total number of pairs).

\begin{figure}[h]
\center{\includegraphics[height=0.24\linewidth]{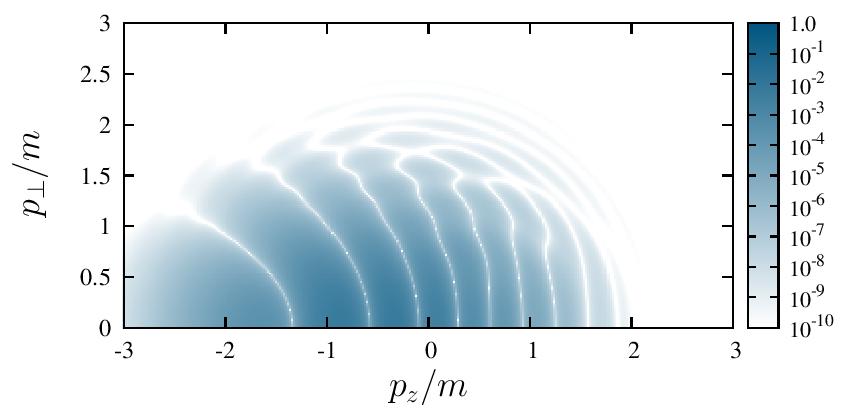}~
\includegraphics[height=0.24\linewidth]{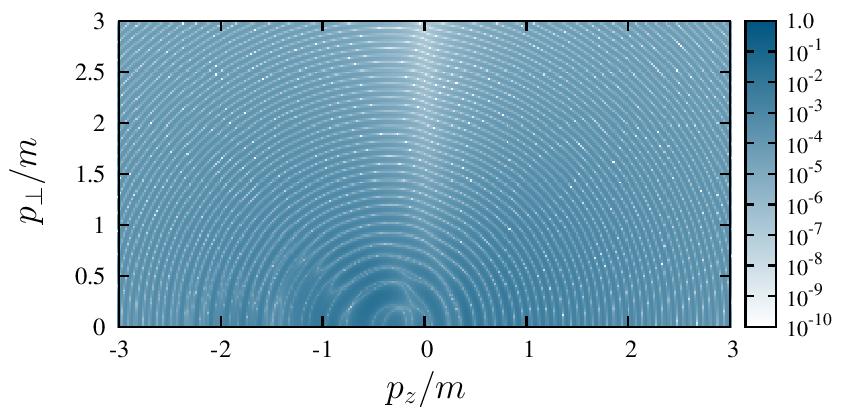}}
\caption{Momentum distribution of electrons created for $\omega = 0.5m$ and both IIa ($N_\text{c} = 2.0$, left) and IIb ($N_\text{c} = 4.0$, right) envelope shapes. The pulse duration is $T\approx 25.1 \, m^{-1}$ (left) and $T\approx 50.3 \, m^{-1}$ (right).}
\label{fig:II_mom_short}
\end{figure}

\indent Again, we make the analogous comparison using smaller values of $N_\text{c}$ (see Fig.~\ref{fig:II_mom_short}).
The role of the pulse shape is crucial now. In the IIb case the pair-production probability oscillates with $p_z$ and $p_\perp$ much faster than in the IIa case. This indicates that the momentum spectrum of electrons created is extremely sensitive to the envelope shape so the distributions presented here and those from Ref.~\cite{hebenstreit_prl_2009} (for the Gaussian pulse shape) are certain to significantly change if one chooses another envelope function (the CEP effects will be discussed in Sec.~\ref{sec:cep}).
%%%
\section{Total number of pairs}\label{sec:total}
\begin{figure}[h]
\center{\includegraphics[height=0.4\linewidth]{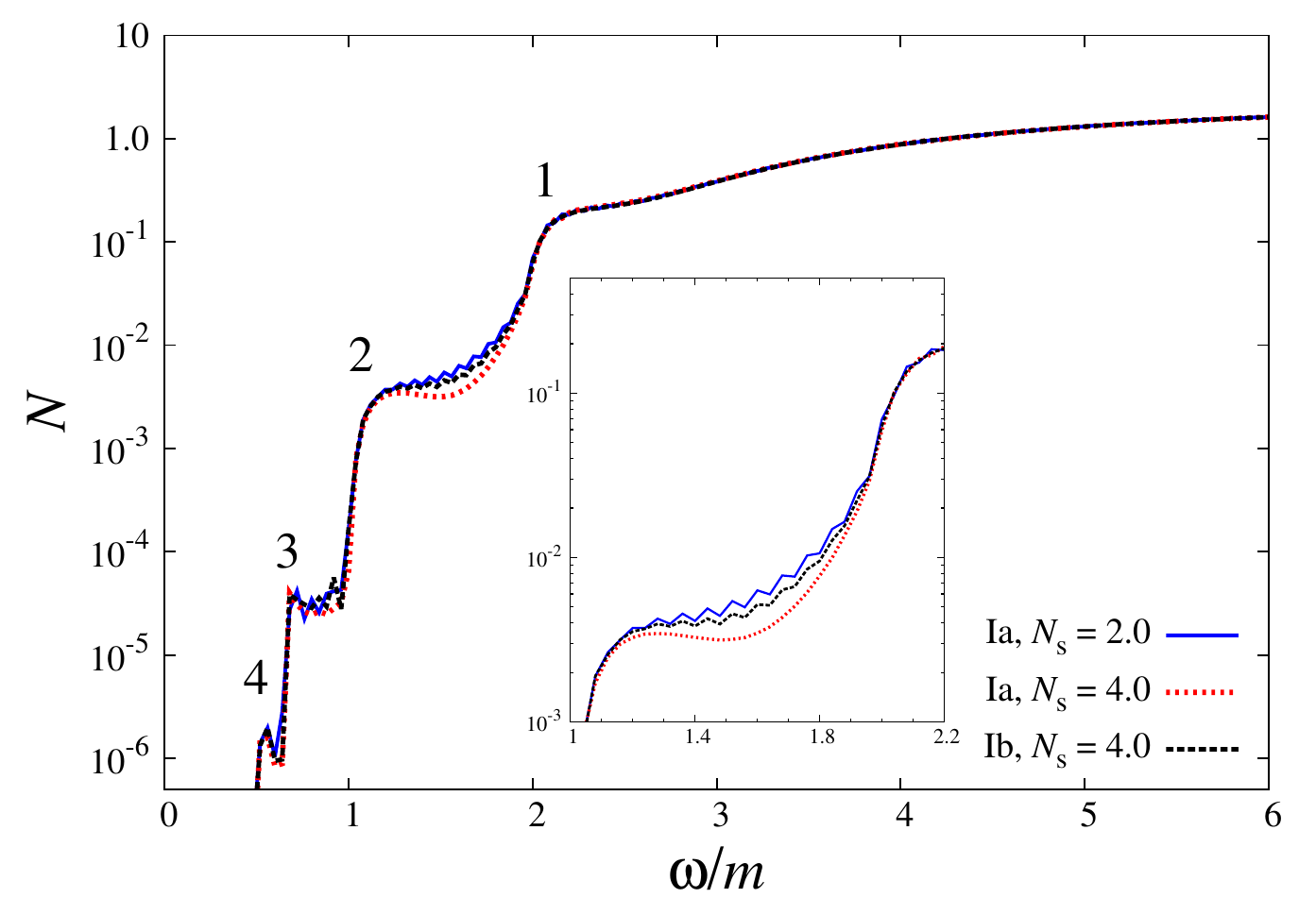}}
\caption{Total number of pairs as a function of $\omega$ for the type-I pulse shapes with different values of $N_\text{s}$ ($T_0 = 50 \pi$).}
\label{fig:total}
\end{figure}

%%%
In order to evaluate the total number of pairs created, one has to perform an integration over the momentum $\boldsymbol{p}$:
\begin{equation}
N = 2 \int \! \mathrm{d}^3 \boldsymbol{p} \, n(\boldsymbol{p}) = 4\pi \int \limits_{-\infty}^{+\infty}\! \mathrm{d} p_z \int \limits_{0}^{+\infty}\! \mathrm{d} p_\perp \, p_\perp n(p_z, p_\perp),
\label{eq:total_integral}
\end{equation}
where the factor $2$ arises due to the spin projection degeneracy. First, we fix $\int F(t) \mathrm{d} t = T_0$ and $E_0 = 0.1 E_\text{c}$ (the value of $\xi$ now varies) for all possible pulse shapes. In Fig.~\ref{fig:total} we compare the total numbers of pairs for three different shapes of type I.
All the curves have large leaps at $\omega_n \approx 2m/n$ which indicate the appearance of additional $n$-photon channels~\cite{abdukerim_plb_2013}. While in the region $\omega \geq 2m$ these three curves coincide, in the plateau regions (e.g., $1.2m \leq \omega \leq 1.9m$) they are different and the total number of pairs is greater for the case of a rapid switching of the external pulse. The oscillations arise since this time we allow the sum $N_\text{c} + N_\text{s}$ to be odd and, therefore, CEP has two possible values --- $\varphi = 0$ and $\varphi = \pi$. When $N_\text{s}$ becomes larger ($N_\text{s} \gtrsim 4$), these oscillations disappear. The role of CEP will be discussed in more detail in the next section.

\begin{figure}[h]
\center{\includegraphics[height=0.345\linewidth]{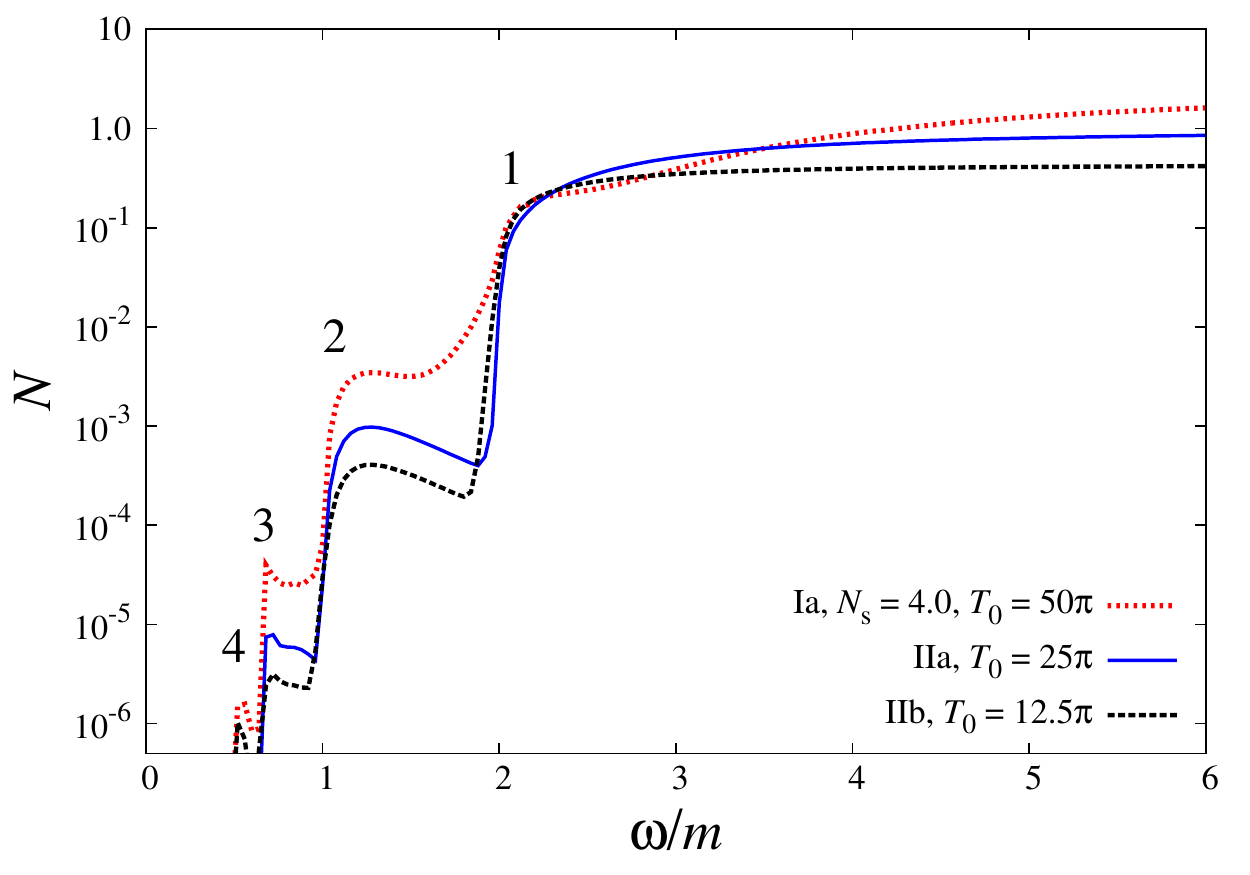}~
\includegraphics[height=0.345\linewidth]{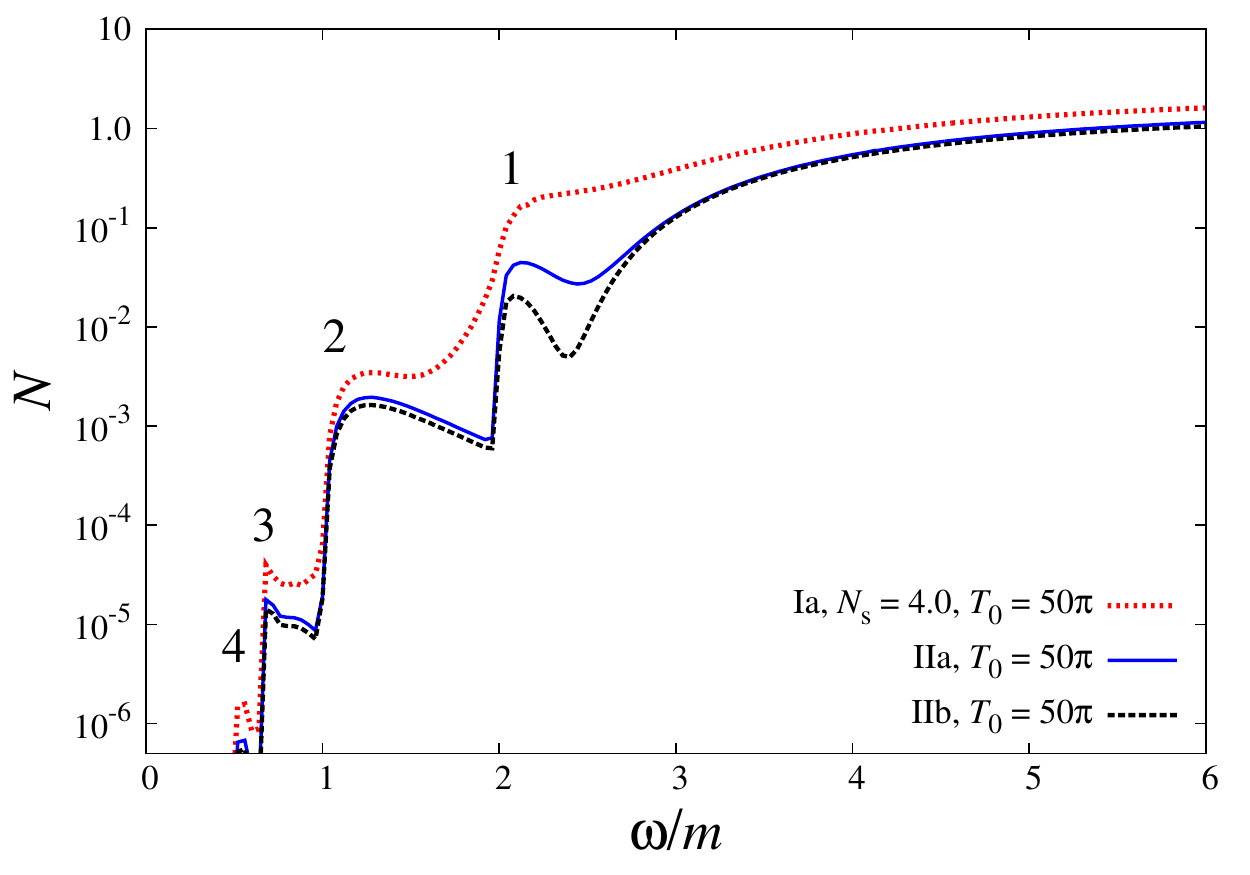}}
\caption{Total number of pairs as a function of $\omega$ for: (left)~type-I and type-II pulses with the same duration, (right)~type-I and type-II pulses with $T_0 = 50\pi$.}
\label{fig:total2}
\end{figure}

\indent In Fig.~\ref{fig:total2} we examine the type-II envelope shapes having the same duration $T$ [Fig.~\ref{fig:total2}(left)] and the same $T_0$ [Fig.~\ref{fig:total2}(right)].
Since for a given $T$ the parameter $T_0$ is greater for the type-I pulses (the choice of $N_\text{s}$, i.e. $\Delta T$, is not very important here), it is no accident that in this case the number of particles produced is larger for almost every $\omega$ [Fig.~\ref{fig:total2}(left)]. Nevertheless, in the region $1.8m \leq \omega \leq 3.6m$ we observe a quite nontrivial behavior. One may expect that these dependences calculated for a given value of $T_0$ should coincide. However, according to Fig.~\ref{fig:total2}(right), the envelopes of the type I for each value of $\omega$ have much more favorable shape. This is in accordance with Ref.~\cite{abdukerim_plb_2013} where it was demonstrated that the ``flat'' super-Gaussian pulse shape is more advantageous than the Gaussian one. In order to maximize the number of pairs created, it is more reasonable to generate a short pulse with rapid switching parts rather than a slowly varying pulse of long duration.
%%%
\section{CEP effects}\label{sec:cep}
\begin{figure}[h]
\center{\includegraphics[height=0.31\linewidth]{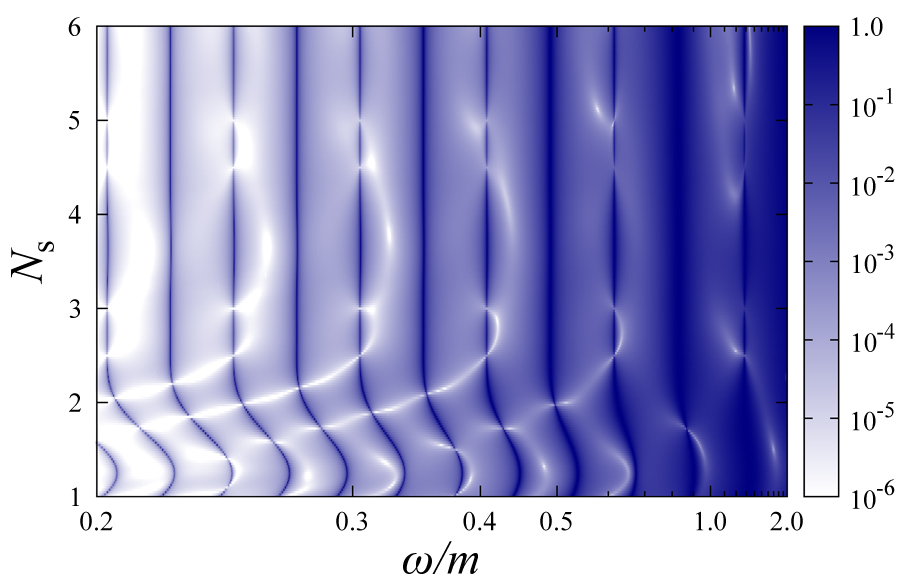}~~
\includegraphics[height=0.31\linewidth]{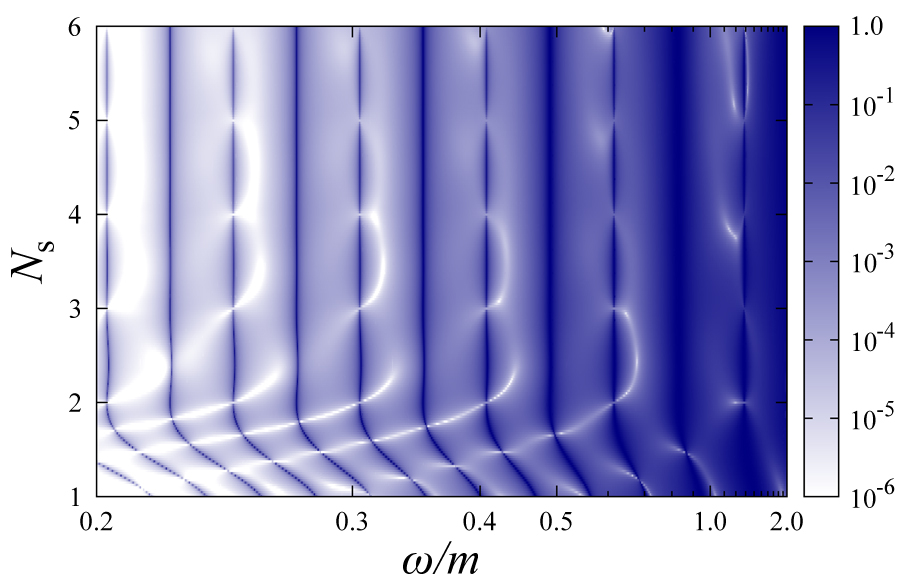}}
\caption{Resonant picture for the Ia envelope shape and $\varphi = \pi/4$ (left) and $\varphi = \pi/2$ (right).}
\label{fig:res_Ia_cep}
\end{figure}

\begin{figure}[h]
\center{\includegraphics[height=0.4\linewidth]{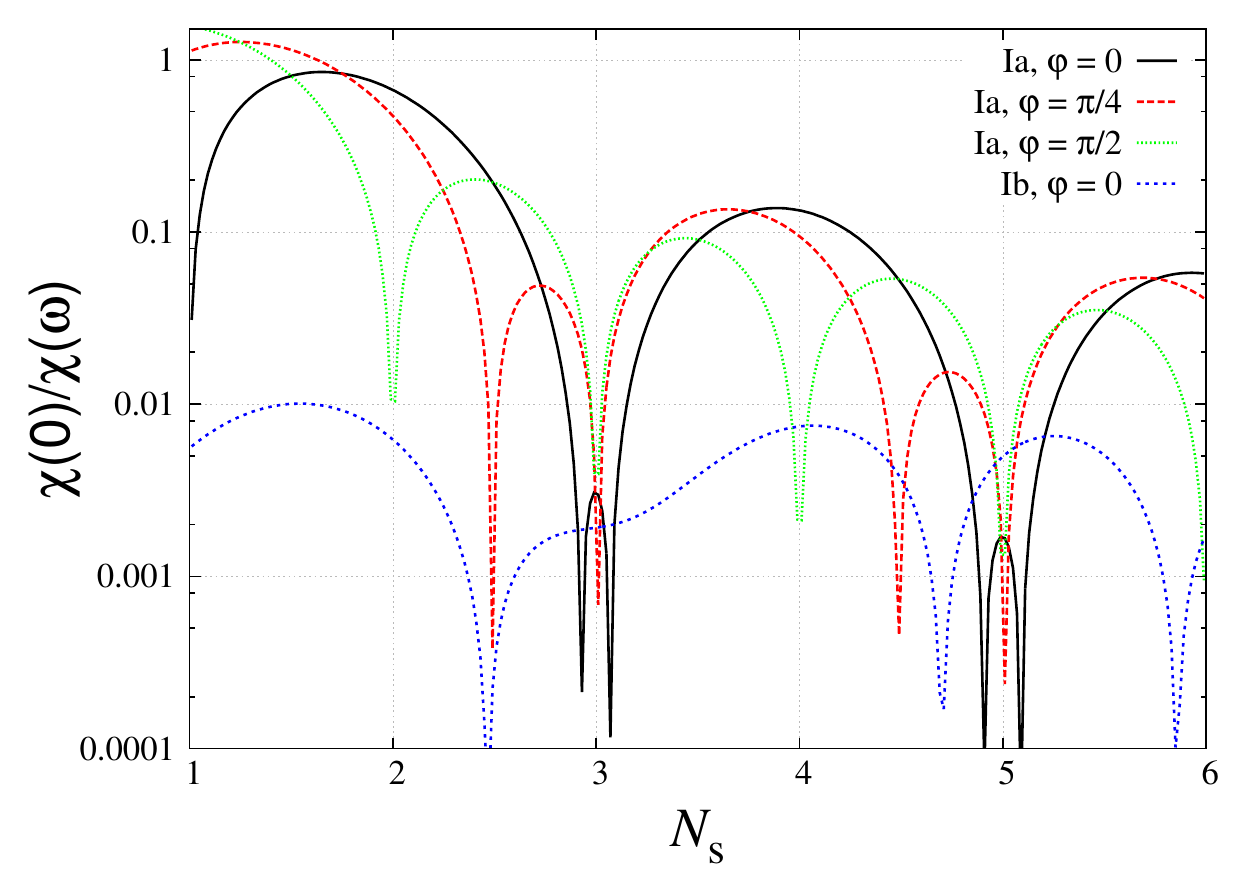}}
\caption{The relative contribution of the Fourier transform of the vector potential at zero energy $k_0 = 0$ for the type-I envelopes with various CEP as a function of $N_\text{s}$.}
\label{fig:fourier}
\end{figure}

%%%
In this section we perform the calculations for arbitrary $\varphi$. Following the same scheme, we start with the discussion of the resonant structure for $\xi = 1.0$. It turns out that the resonant pattern in the case of the Ib envelope shape remains the same no matter which value of CEP is chosen. Nevertheless, in the Ia case for $N_\text{s} \lesssim 2.5$ the peaks are shifted differently for different $\varphi$. In Fig.~\ref{fig:res_Ia_cep} this is depicted for $\varphi = \pi/4$ (left) and $\varphi = \pi/2$ (right).
When $N_\text{s}$ becomes greater ($N_\text{s} \gtrsim 2.5$), the positions of the resonances are similar for all $\varphi$. However, we observe that the values of $N_\text{s}$ for which the even resonances disappear depend on $\varphi$. In addition to those given by $N_\text{s} = 2k + 1$ ($k \in \mathbb{N}$) we also have $N_\text{s} = 2k + 1 - 2\varphi/\pi$. This can be accounted for by the Fourier analysis of the external time-dependent pulses. Let us introduce the function
\begin{equation}
\chi (k_0) = \sqrt{f^2 (k_0) + g^2 (k_0)},
\label{eq:fourier_chi}
\end{equation}
where the functions
\begin{eqnarray}
f(k_0) &=& \int \limits_{t_\text{in}}^{t_\text{out}} \mathrm{d} t \cos k_0 t \, \big [A_z(t) - A_z (t_\text{out}) \big ], \label{eq:fourier_f}\\
g(k_0) &=& \int \limits_{t_\text{in}}^{t_\text{out}} \mathrm{d} t \sin k_0 t \, \big [A_z(t) - A_z (t_\text{out}) \big ] \label{eq:fourier_g}
\end{eqnarray}
represent the real and imaginary parts of the Fourier transform of the vector potential, respectively. The function $\chi (k_0)$ is gauge invariant and continuous (we discard the singular contributions that arise from the nonvanishing parts outside the interval $[t_\text{in},~t_\text{out}]$). It turns out that this function has two pronounced peaks at $k_0 = \pm \omega$ and may also be large at $k_0 = 0$ for some special parameters of the external pulse. At other points $k_0 \in \mathbb{R}$ it is several orders of magnitude smaller. In Fig.~\ref{fig:fourier} we present the ratio $\chi (0)/\chi (\omega)$ as a function of $N_\text{s}$ for the type-I pulses with different values of CEP.
One observes that the ``zero-energy'' contribution for the Ia case is much larger than that for the Ib envelope shape. However, for several values of $N_\text{s}$ it also becomes insignificant (we observe relatively sharp minima at $N_\text{s} = 2k + 1$ and $N_\text{s} = 2k + 1 - 2\varphi/\pi$, where $k \in \mathbb{N}$). These values exactly correspond to those for which the resonant peaks vanish [see Figs.~\ref{fig:res_I_to}(left) and \ref{fig:res_Ia_cep}]. Thus, we conclude that the appearance of the even resonances is due to the nonmonochromaticity of the external field which effectively violates the selection rule discussed previously. Since the zero-energy mode in the Ib case is small for all values of $N_\text{s}$, the even resonances in Fig.~\ref{fig:res_I_to}(right) are always strongly suppressed (the same holds also true for the type-II envelopes).

\begin{figure}[h]
\center{\includegraphics[height=0.243\linewidth]{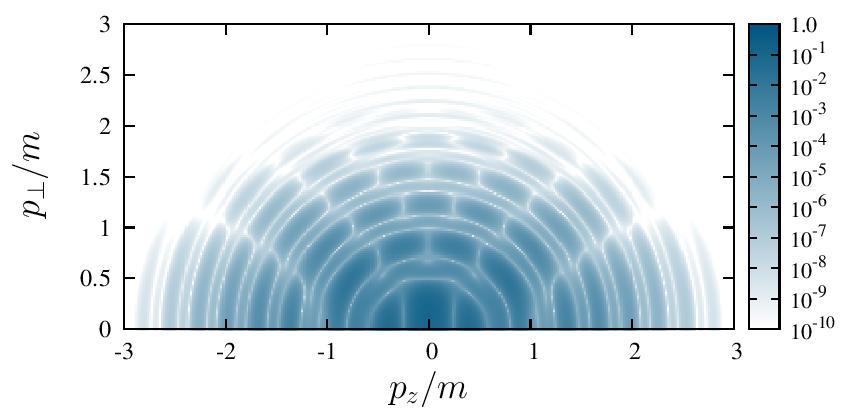}~
\includegraphics[height=0.243\linewidth]{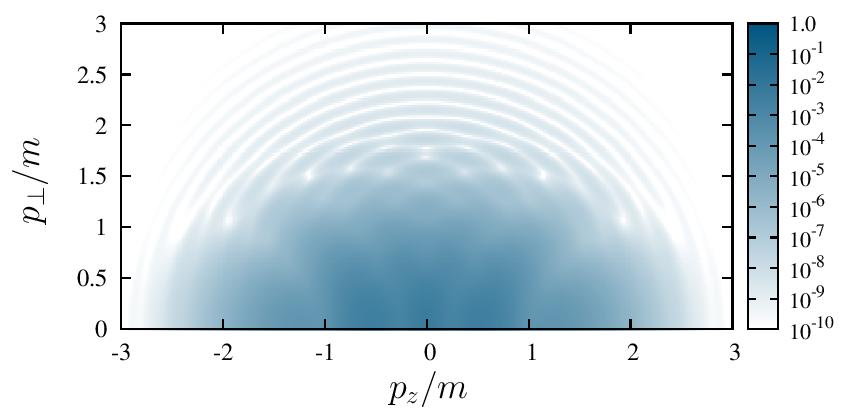}}
\caption{Momentum distribution of electrons created for the Ia pulse with $N_\text{c} = 2.0$ and $N_\text{s} = 4.0$ (left) and the IIa pulse with $N_\text{c} = 2.0$ (right). For both figures $\varphi = \pi/2$ and $\omega = 0.5m$.}
\label{fig:mom_cep}
\end{figure}

\indent Another important feature relates to the momentum spectrum of electrons created. As was stated previously, the pulse shape does not play any important role when the number of cycles $N_\text{c}$ is large. A similar conclusion can be drawn regarding CEP. However, for small $N_\text{c}$ the CEP effects can be easily revealed. In Fig.~\ref{fig:mom_cep} we present the momentum distributions for the Ia and IIa pulse shapes with $\varphi = \pi/2$.
Comparing these results with those displayed in Figs.~\ref{fig:I_mom_short} and \ref{fig:II_mom_short}, we note that CEP may substantially alter the momentum spectrum, especially, in the case of the type-II envelopes. In Ref.~\cite{hebenstreit_prl_2009} it was shown that when the electric pulse with the Gaussian envelope is antisymmetric, i.e. it obeys $E(-t) = -E(t)$ [according to Eq.~(\ref{eq:field_gen}), it corresponds to $\varphi = 0$], then the pair-production probability oscillates and vanishes at its minima (it is clearly seen in Fig.~\ref{fig:II_mom_short} for the type-II pulses). When the field becomes symmetric ($\varphi = \pi/2$), these oscillations disappear~\cite{hebenstreit_prl_2009} as can be observed in Fig.~\ref{fig:mom_cep}(right). However, the pulses of the type I do not possess this feature.

\indent Finally, we note that we have not found any significant CEP effects on the total number of pairs. Whereas this parameter may be responsible for various very interesting signatures regarding differential characteristics of the pair-production process, it appears not to have a notable impact on the total number of particles created.
%%%
\section{Discussion and conclusion}\label{sec:discussion}
Within the present study we examined four different pulse shapes for their various parameters. In order to provide a comprehensive analysis, we considered several characteristics of the pair-production process: the resonant structure, Rabi oscillations, momentum spectrum of particles created, and total number of pairs. Besides, we studied two different types of the envelope shape (two specific forms for each type) and the role of CEP, which allowed us to conduct an extensive study. We believe that the present investigation considerably expands the previous findings and further clarifies the role of the pulse shape.

\indent Our main results can be summarized as follows:
\begin{itemize}
\item The resonant pattern for $\boldsymbol{p} = 0$ strongly depends on the shape of the temporal profile. In the case of the Ib envelopes and the type-II pulses only the odd resonances occur and the resonant structure is almost independent of $N_\text{s}$, while, in contrast to the results of previous studies, in the Ia case one can observe the even resonances and the picture is different for different $N_\text{s}$ in the region $N_\text{s} \lesssim 2.5$. Moreover, it was found that the peaks with even $n$ disappear for a discrete set of specific values of $N_\text{s}$ that depend on CEP. This nontrivial property manifests itself due to the fact that the laser pulse is not monochromatic. The interference among modes having different frequencies could considerably change the resonant picture.
\item The Rabi frequencies may be very sensitive to the pulse shape, especially, in the case of the type-II envelopes. Besides, it was found that the Rabi frequencies are much larger for the type-I case. 
\item The resonant structure of the momentum distribution of particles created is different for the type-I and type-II envelope shapes. Nevertheless, when the external pulse contains a large number of cycles, the distribution is almost independent of the form of the type-I switching function and of the type-II pulse shape. However, for the case of a small number of cycles all the parameters defining the pulse shape (including CEP) have a great influence on the momentum spectrum.
\item The total number of pairs produced strongly depends on the envelope shape. For the type-I pulses it is much greater than for the type-II case. Envelopes with rapid switching functions are more advantageous than slowly-varying pulse profiles.
\end{itemize}

\indent In our calculations we employed large values of $\omega$ (and, therefore, large $E_0$) for two reasons. First, this allowed us to study the resonant pattern of the $n$-photon peaks as well as the resonant Rabi oscillations. Second, large values of the laser frequency help one to avoid very time-consuming computations, which is particularly important for the calculations of the total number of pairs created. Although the corresponding parameters of laser fields are not currently available in the experiment, we suppose that most features revealed within the present study (listed above) are likely to be valid for weaker fields and lower frequencies as well.

\indent In addition, we point out that the pulse shape effects may be significant in the context of dynamically assistant pair production~\cite{schutzhold_prl_2008} (see also Refs.~\cite{akal_prd_2014, otto_plb_2015} and references therein). This was examined in Refs.~\cite{linder_prd_2015, hebenstreit_plb_2014, nuriman_plb_2012, kohlfuerst_prd_2013}. Further analysis of these effects may be a very interesting and important prospect for future studies.

\indent Finally, one has to note that spatial variations of the external field could play a significant role~\cite{hebenstreit_prl_2011, hebenstreit_prd_2010, ruf_prl_2009, woellert_prd_2015, aleksandrov_prd_2016}. Although taking them into account appears to be a very difficult task, one of the most important subjects of our future investigations is the application of the technique described in Ref.~\cite{aleksandrov_prd_2016} to the corresponding multidimensional scenarios involving strong laser fields.
%%%
\section*{Acknowledgments}
This investigation was supported by RFBR (Grant No.~16-02-00334) and by Saint Petersburg State University (SPbU) (Grants No. 11.65.41.2017, No. 11.42.987.2016, No. 11.42.939.2016, and No. 11.38.237.2015). I. A. A. acknowledges the support from the German-Russian Interdisciplinary Science Center (G-RISC) funded by the German Federal Foreign Office via the German Academic Exchange Service (DAAD), from TU Dresden (DAAD-Programm Ostpartnerschaften), and from the Supercomputing Center of Lomonosov Moscow State University.
%%%
\appendix
\section{Numerical approach}\label{sec:appendix}
We assume that the external field vanishes for $t \leq t_\text{in}$ and for $t \geq t_\text{out}$ \big[e.g., for the Ia envelope defined by Eq.~(\ref{eq:F_Ia}), we choose $t_\text{in} = - t_\text{out} = -T/2 - \Delta T$\big]. Following the rigorous QFT formalism described in Ref.~\cite{fradkin_gitman_shvartsman}, we solve the Dirac equation in order to construct two orthonormal and complete sets of time-dependent solutions with given asymptotic behavior:
\begin{equation}
{}_\zeta \Psi_n (t_\text{in}, \boldsymbol{x}) = {}_\zeta \Psi^{(0)}_n (\boldsymbol{x}),\quad {}^\zeta \Psi_n (t_\text{out}, \boldsymbol{x}) = {}^\zeta \Psi^{(0)}_n (\boldsymbol{x}),\label{eq:psi_in_out}
\end{equation}
where ${}_\zeta \Psi^{(0)}_n (\boldsymbol{x})$ and ${}^\zeta \Psi^{(0)}_n (\boldsymbol{x})$ are the eigenfunctions of the Dirac Hamiltonian considered at $t=t_\text{in}$ and $t=t_\text{out}$, respectively, and $\zeta$ is the sign of the corresponding (energy) eigenvalues. It turns out that these sets of solutions contain all the information about the quantities to be analyzed within the present study. For instance, the mean number of electrons (positrons) created with the given quantum numbers $m$ can be evaluated via~\cite{fradkin_gitman_shvartsman}
\begin{eqnarray}
&&n^-_m = \sum_n G({}^+|{}_-)_{mn} G({}_-|{}^+)_{nm}, \label{eq:num_el}\\
&&n^+_m = \sum_n G({}^-|{}_+)_{mn} G({}_+|{}^-)_{nm}, \label{eq:num_pos}
\end{eqnarray}
where the $G$ matrices can be defined as the following inner products
\begin{eqnarray}
&&G({}^\zeta|{}_\kappa)_{nm} = ({}^\zeta \Psi_n,~{}_\kappa \Psi_m), \label{eq:G_inner_product_1}\\
&&G({}_\zeta|{}^\kappa)_{nm} = ({}_\zeta \Psi_n,~{}^\kappa \Psi_m). \label{eq:G_inner_product_2}
\end{eqnarray}

\indent The Dirac equation in the presence of an external background reads
\begin{equation}
\big ( \gamma^\mu \big [i \partial_\mu - e A_\mu (t, \boldsymbol{x}) \big ] - m \big) \Psi (t, \boldsymbol{x}) = 0.
\label{eq:dirac_general}
\end{equation}
We use the conventional substitution $\Psi = \big [ \gamma^\mu \big (i \partial_\mu - e A_\mu \big ) + m \big] \psi$ which yields (see, e.g., Refs.~\cite{fradkin_gitman_shvartsman, gav_git_prd_1996})
\begin{equation}
\big ( [i \partial - e A ]^2 - m^2 - \frac{ie}{2} \gamma^\mu \gamma^\nu F_{\mu \nu} \big ) \psi (t, \boldsymbol{x}) = 0,
\label{eq:dirac_squared}
\end{equation}
where $F_{\mu \nu} = \partial_\mu A_\nu - \partial_\nu A_\mu$. Within DA one can choose
\begin{equation}
A_0 = 0, \quad A_x = A_y = 0,\quad A_z (t) = - \int \limits^t E(t') \mathrm{d} t'.
\label{eq:gauge_t}
\end{equation}
Hence, Eq.~(\ref{eq:dirac_squared}) takes the form
\begin{equation}
\big [\partial_t^2 - \Delta + 2 ie A_z (t) \partial_z + e^2 A_z^2 (t) + m^2 + ie \gamma^0 \gamma^3 E (t) \big ] \psi (t, \boldsymbol{x}) = 0,
\label{eq:dirac_t}
\end{equation}
which can be reduced to a scalar equation if one represents the function $\psi (t, \boldsymbol{x})$ as
\begin{equation}
\psi_n (t, \boldsymbol{x}) = \psi_{\boldsymbol{p}, s, r} (t, \boldsymbol{x}) = \mathrm{e}^{i\boldsymbol{p} \boldsymbol{x}} v_{s,r} \varphi_{\boldsymbol{p}, s, r} (t),
\label{eq:psi_t_spin}
\end{equation}
where $v_{s,r}$ ($s = \pm 1$, $r = \pm 1$) are the constant orthonormal eigenvectors of the matrix $\gamma^0 \gamma^3 = \alpha^3$. The scalar function $\varphi_{\boldsymbol{p}, s, r} (t)$ is a solution of the following equation:
\begin{equation}
\big [\partial_t^2 + (p_z - e A_z (t))^2 + p_\perp^2 + m^2 + iesE(t) \big ] \varphi_{\boldsymbol{p}, s, r} (t) = 0,\quad \boldsymbol{p} = (\boldsymbol{p}_\perp, p_z).
\label{eq:dirac_t_phi}
\end{equation}
By solving this ordinary differential equation, we obtain the matrix elements of the $G$ matrices and employ Eqs.~(\ref{eq:num_el}) and (\ref{eq:num_pos}) in order to calculate the spectrum of particles produced. Since the generalized momentum is conserved, the matrices are diagonal: $G ({}^\kappa|{}_\zeta)_{mn} = \delta_{mn} \, g ({}^\kappa|{}_\zeta)_{n}$. Note that due to this relation, in the region $t \leq t_\text{in}$ ($t \geq t_\text{out}$) each out (in) solution with the sign $\zeta$ represents as a linear combination of the only two in (out) solutions with $\zeta$ and $-\zeta$, respectively. The corresponding coefficients contain the necessary matrix elements and can be evaluated by using the function $\varphi$ and its derivative at the point $t = t_\text{in}$ ($t = t_\text{out}$). Thus, it is not necessary to perform the integration according to Eqs.~(\ref{eq:G_inner_product_1}) and (\ref{eq:G_inner_product_2}).
%%%
%%%%%

\end{document}